\documentclass[letter]{aa}

\usepackage{graphicx}
	\graphicspath{{./img/},{./}}
\usepackage{txfonts}
\usepackage{hyperref}
\usepackage{textcomp,mathabx}
\usepackage{siunitx}
    \DeclareSIUnit{\astronomicalunit}{AU}
	\DeclareSIUnit{\parsec}{pc}
	\DeclareSIUnit{\earthmass}{M_\Earth}
	\DeclareSIUnit{\solarmass}{M_\Sun}
	\DeclareSIUnit{\jupitermass}{M_J}
	\DeclareSIUnit{\year}{yr}
\usepackage{CJKutf8}
\usepackage{placeins}

\begin{document} 

    \bibliographystyle{aa}
    \title{Late accretion offers pathway to misaligned disk around\\ the planet-hosting IRAS~04125+2902}

    \author{L.-A. H\"uhn\inst{1}
        \and
        H.-C. Jiang \begin{CJK*}{UTF8}{gbsn}(蒋昊昌)\end{CJK*}\inst{2}
        \and
        C. P. Dullemond\inst{1}
        }

    \institute{Institut für Theoretische Astrophysik, Zentrum für Astronomie der Universität Heidelberg, Albert-Ueberle-Str. 2, 69120 Heidelberg, Germany\\
    \email{\href{mailto:huehn@uni-heidelberg.de}{huehn@uni-heidelberg.de}}
    \and
    Max-Planck Institute for Astronomy (MPIA), Königstuhl 17, 69117 Heidelberg, Germany
    }

    \date{\today}

    \abstract
    {We present a 3D hydrodynamical simulation of the accretion of a gas cloudlet onto the IRAS~04125+2902 binary system, where the 3-Myr primary hosts a transiting planet. We demonstrate that such an accretion event can naturally produce a circumstellar disk that is misaligned with respect to the rest of the system, consistent with the observed misaligned transition disk. In the model, the prescribed orbital plane of the cloudlet is largely retained by the resulting circumstellar disk after undergoing gravitational interactions with the secondary during the initial accretion. After ${\sim}4.4$ binary orbits, a disk with $R_d=\SI{300}{\astronomicalunit}$ has formed around the stellar primary made of ${\sim}13\%$ of the cloudlet mass, $M_\mathrm{d,p}=\SI{2.1e-3}{\solarmass}$. The companion also retains some of the cloudlet's mass and forms a disk with $M_\mathrm{d,c}=\SI{9.3e-5}{\solarmass}$, though only the transition disk around the primary has been observed. Our findings highlight the importance of considering mass inflow onto protoplanetary disk for their evolution.}


    \maketitle

\section{Introduction}
In the current picture of exoplanet formation, planets are the products of multiscale processes whereby initially micron-sized dust grains grow into planetary embryos that migrate and accrete (for a review, see \citealt{drazkowska2023}). These processes occur in the protoplanetary disk around a young star, and as a result, fully formed planets are generally expected to lie in the same orbital plane as the disk out of which it formed.

Recent observations of the IRAS~04125+2902 binary system, located inside the Taurus star-forming region, reveal a $\SI{0.3}{\jupitermass}$ transiting giant planet \citep{barber2024} around the stellar primary, with a semi-major axis of \SI{0.077}{\astronomicalunit}. This discovery paints a peculiar picture of the system configuration. While it appears that the orbital axis of the giant planet is consistent with the orbital axis of the ${\sim}4^{\prime\prime}$ companion, the transition disk observed in this system is misaligned relative to both. The stellar inclination is $i_\star > \SI{78}{\degree}$, consistent with the planetary orbit, and the inclination of the binary orbit is $i_c = 94.5^{+10.9}_{-4.7}$ degrees with respect to the plane of the sky. In contrast to the nearly edge-on configuration of all system components, the inclination of the transition disk is more face-on, with $i_d=31\pm \SI{12}{\degree}$, firstly estimated based on low-resolution continuum data \citep{espaillat2015} from the Submillimeter Array (SMA), and then confirmed by CO emission and high-resolution continuum from the Atacama Large Millimeter/submillimeter Array (ALMA) \citep[see Appendix~\ref{sec:app_alma}]{bosschaart2025,shoshi2025}.

A transition disk that is highly misaligned with other system components, especially with respect to the giant planet's orbit, is unexpected. Planets embedded in protoplanetary disks are subject to eccentricity and inclination damping due to planet-disk interactions \citep{tw04,mn09,bk11}, suggesting the misalignment would have to arise by gravitational interaction after the natal inner disk has dissipated. Alternatively, as the alignment of the stellar spin and binary orbit suggests, the entire disk could have been initially aligned with the planet's current orbit, but become misaligned later. The interaction with the known companion would preferentially align the disk to the binary orbit \citep{christian2022}. However, we note that, while the inclination of the planet and binary orbit are consistent, the planet's position angle is not constrained. In general, disk-planet misalignment could, for example, be caused by an outer companion \citep{papaloizou1995,zl18}, but the secondary star and giant planet are the only detected massive objects in IRAS~04125+2902. A stellar flyby could also cause disk misalignment \citep{cp93,nealon2020}. In fact, for the system at hand, it has been suggested that the system was initially an unstable triplet, and that the ejection of the third star induced the misalignment \citep{nealon2025}. However, such a scenario implies further misalignment of other system components and offers an unlikely explanation if the binary and planetary orbits are truly aligned.

We present an alternative explanation for the misalignment. Using 3D hydrodynamical simulations, we show that the observed transition disk could be a second-generation disk accreted during an encounter with a gas cloudlet. Previous work has demonstrated that an encounter with a cloudlet can lead to the formation of a second generation disk around single stars \citep{dullemond2019,kuffmeier2020}, and late infall events have been proposed to explain streamer structures onto protoplanetary disks (e.g., \citealt{ginski2021,hanawa2024,calcino2025,speedie2025}). In our simulation, a misaligned disk arises naturally in a similar fashion from the accretion of the gas cloudlet, and an initial misalignment between the secondary and the cloudlet orbit can survive despite the gravitational influence of the companion and mass transfer events. This suggests that the observed transition disk need not be the same disk that formed the giant planet, and no additional massive object is required to explain its misalignment.

\section{Methods}\label{sec:methods}
We perform 3D hydrodynamical simulations using the \texttt{FARGO3D} code \citep{bm16}, a grid-based code which utilizes the FARGO orbital advection algorithm \citep{masset2000}. The simulations use a spherical grid centered on the primary star that is spaced logarithmically in the radial dimension, extending from $\theta_\mathrm{min}=\SI{20}{\degree}$ to $\theta_\mathrm{max}=\SI{160}{\degree}$, and from $r_\mathrm{min}=\SI{10}{\astronomicalunit}$ to $r_\mathrm{max}=\SI{5000}{\astronomicalunit}$. We choose the number of grid cells to be $N_r=280$, $N_\theta=110$ and $N_\phi=280$. The resulting resolution is $\Delta\phi=\Delta\theta=\SI{1.3}{\degree}$ and $\Delta r/r=0.022\simeq\Delta\phi$, corresponding to 2 cells per pressure scale height (cps) at \SI{10}{\astronomicalunit} (3 at \SI{50}{\astronomicalunit}). The system is integrated up to a physical time of $t_\mathrm{end}=\SI{94.8}{\kilo\year}$, corresponding to $4.4$ binary orbits. Both the polar and radial boundaries are open, so that mass can leave the simulation domain, but cannot enter.

The simulations assume a locally isothermal equation of state, $P=\rho c_s^2$, where $P$ is the gas pressure, $\rho$ is the density and $c_s$ is the radius-dependent isothermal sound speed. For the initial condition, no disk is around either star. The respective temperature is determined solely by passive irradiation of the central star with a floor value $T_\mathrm{floor}$, so that the aspect ratio $h=H/r$ of the eventually accreted disk is given by $h = h_0(r/r_0)^\beta$, with $H$ being the gas pressure scale height, the radial distance $r$, constants $r_0$ and $h_0$ and the flaring index $\beta$. We use $T_\mathrm{floor}=\SI{10}{\kelvin}$, so that $c_{s,\mathrm{floor}}=\SI{0.19}{\kilo\meter\per\second}$, reached at $r_\mathrm{floor}=\SI{796}{\astronomicalunit}$. We adopt $h_0=0.038$, $r_0=\SI{5.2}{\astronomicalunit}$ and $\beta=0.25$. A kinematic viscosity $\nu$ is included, given as a constant $\alpha$ parameter \citep{ss1973} of $\alpha=\num{e-3}$, so that $\nu=\alpha c_s^2/\Omega$. The kinematic viscosity is included in every cell in the simulation grid, and calculated with respect to the primary even if material is not bound to it, so that $\Omega=\sqrt{GM_\star/r^3}$ with $M_\star=\SI{0.7}{\solarmass}$ \citep{barber2024}. As a result, the viscosity of the cloudlet is negligible. We introduce a numerical background density of $\rho_\mathrm{bg}=\SI{e-21}{\gram\per\centi\meter\cubed}$, comparable with the typical density in Taurus \citep{pineda2010}. We neglect the small gravitational influence of the giant planet in our model.

It may seem natural to put the origin of the coordinate system on the barycenter. However, for simplicity, for example, in the implementation of the boundary conditions, and numerical reasons, such as numerical diffusion due to the movement of the inner disk edge in a barycentric system, the \texttt{FARGO3D} code centers the coordinate system on the central star. The inertial force caused by the acceleration of the star by the companion -- and therefore the acceleration of the entire coordinate system -- is accounted for in the gas potential. This is commonly referred to as the ``indirect term'' \citep{crida2025}. The gravitational forces and corresponding frame acceleration by the gas onto the two stars is, however, not included.

\subsection{Binary companion}
For the binary orbit, we employ the semi-major axis determined by orbital fitting, $a=4.66^{\prime\prime}$ \citep[][Extended Data Fig. 5]{barber2024}, resulting in $a\approx\SI{742}{\astronomicalunit}$, and a mass of $M_\mathrm{s}=\SI{0.17}{\solarmass}$. As the orbital eccentricity of the companion is unconstrained, we choose $e=0$ for simplicity. We choose a coordinate system such that the companion has an inclination of $i_c=\SI{60}{\degree}$, whereas the cloudlet orbit is not inclined, that is, it lies in the x-y plane. The companion is initialized at the apoastron. To model the orbital motion, we use the 5th order Runge-Kutta N-body solver included in \texttt{FARGO3D}.

\subsection{Cloudlet initial condition}
The gas cloudlet is initially modeled as a sphere with radius $R_\mathrm{c}=\SI{1500}{\astronomicalunit}$. The center of the sphere is placed on a hyperbolic orbit with impact parameter regarding the center of mass of $b=0.52G(M_\star+M_\mathrm{s})/v_\infty^2=\SI{1615}{\astronomicalunit}$, with a speed at infinity of $v_\infty=\SI{0.5}{\kilo\meter\per\second}$ and at a distance of $r_\mathrm{ini}=\SI{3000}{\astronomicalunit}$, resulting in a lapse time to the periapsis of \SI{13.4}{\kilo\year}. This implies an eccentricity of $e=1.43$ and a periapsis distance of $r_\mathrm{p}=\SI{393}{\astronomicalunit}$. The initial velocity is assumed to be constant across the extent of the cloudlet and given by the orbital velocity. The total mass of the cloudlet is $M_\mathrm{c}=\SI{0.016}{\solarmass}$ with an initially uniform density of $\rho_\mathrm{c}=\SI{6.72e-19}{\gram\per\centi\meter\cubed}$, a value comparable to the density of a filamentary structure in Taurus \citep{palmeirim2013}.

\section{Results}\label{sec:results}
We find that the gas cloudlet is accreted and forms circumstellar disks around both the primary and the secondary.\footnote{For a video, see \url{https://leonhuehn.de/files/iras04125.mp4}} As the cloudlet and binary system undergo gravitational interactions, the orbital misalignment evolves over time from the initial value of $\vartheta_\mathrm{d,o} = \SI{60}{\degree}$. Due to the steady infall of material with different angular momentum, caused by the fall-back of gas surrounding the system after the initial encounter, the disk also becomes warped. However, after ${\sim}4.4$ binary orbits, we find that the misalignment between disk and the binary orbit is still significant, that is, the gravitational interactions with the binary did not force it on an aligned orbit to a significant extent.

\begin{figure}[htp]
    \centering\includegraphics[width=.9\linewidth]{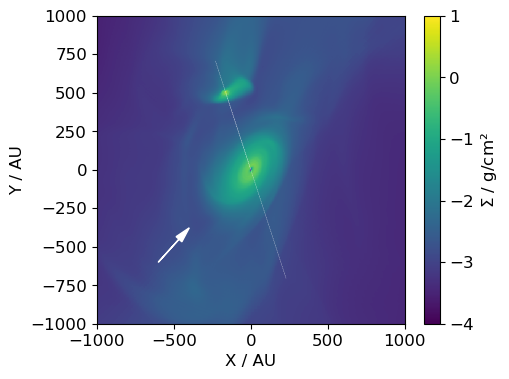}
    \caption{Gas column density at $t_\mathrm{end}=\SI{94.8}{\kilo\year}$. The camera angle is such that the position angle of the binary orbit corresponds to the observed value. The orbit is indicated by the white dotted line, and the white arrows indicate the initial infall direction.}
    \label{fig:figure_1}
\end{figure}%
The final system configuration is shown in Fig. \ref{fig:figure_1}, depicting the column density of the gas in the "true" line of sight to the source, matching the known position angle of the binary orbit, $\mathrm{PA}\approx \SI{198}{\degree}$ \citep{barber2024}. Qualitatively, we find that the overall shape of the primary disk matches that of recent ALMA observations, presented in Appendix~\ref{sec:app_alma}, though the simulated disk is larger than observed. The misalignment with the binary orbit is apparent as the view is edge-on to the binary orbit. While Fig. \ref{fig:figure_1} shows that there is still material left over from the accretion event, the system is dynamically quiescent at this stage, and the gas remnants are expected to diffuse away or get accreted on longer time scales. We show more simulation snapshots in Appendix \ref{sec:app_time_series}.

\begin{figure}[htp]
    \centering\includegraphics[width=.9\linewidth]{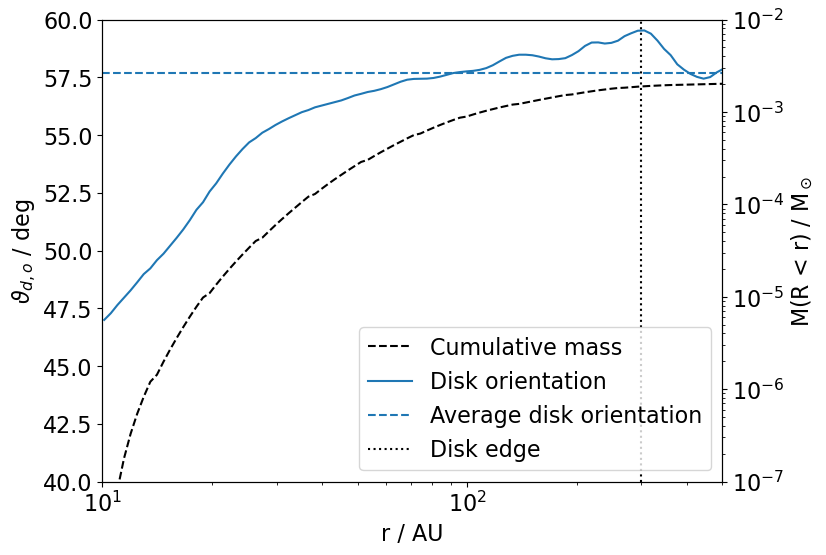}
    \caption{Orientation of the primary disk. The angle between the angular momentum vector of a ring at the radius indicated on the horizontal axis and the binary orbit is shown as a solid blue line. The mass-weighted average value is given by the dashed blue line. For reference, the disk's cumulative mass is indicated by the dashed black line, whereas the dotted vertical black line denotes the limiting radius for the mass-weighting.}
    \label{fig:figure_3}
\end{figure}%
Figure \ref{fig:figure_3} shows the final orientation of the disk around the primary, that is, the angle between the mass-weighted average angular momentum vector of a spherical shell at a given radius and the angular momentum vector of the binary orbit. For reference, it also shows the cumulative mass of the primary disk as a function of radius and a rough estimate for the disk extent, $R_\mathrm{disk}\approx\SI{300}{\astronomicalunit}$. More information on the disk masses is provided in Appendix \ref{sec:app_disk_mass}. We find a misalignment varying between ${\sim}\SI{47}{\degree}$ and ${\sim}\SI{60}{\degree}$ due to the warp (see above). The overall disk misalignment, that is, the mass-weighted average of all spherical shell orientations, is $\bar\vartheta_\mathrm{d,o}\approx\SI{58}{\degree}$. This value is consistent with the observed inclination within $1\sigma$, under the assumption that the position angle of the transition disk is close to zero, which is subject to observational uncertainty. We note that on long time scales, the misalignment could increase or decrease due to von Zeipel-Kozai-Lidov oscillations, which the disk is susceptible to due to its large extent \citep{zl17,nealon2025}.

Our proposed scenario of a second-generation disk, which undergoes further evolution toward a transition disk after the end of the simulation, being the explanation for the disk-planet misalignment, presented in greater detail in Appendix \ref{sec:app_formation_scenario}, has two defining features that set it apart from flyby-based scenarios. First, the accretion event naturally leads to the formation of disks around not only the primary, but both stars. Second, the accretion event does not perturb any other elements of the overall system. In other words, while explanation based on gravitational interaction have difficulties explaining the misalignment of the disk if all other elements are truly aligned with each other, this is the natural and expected outcome of a late accretion event.

\section{Discussion}\label{sec:discussion}
\subsection{Model considerations}
As a proof-of-concept approach, we realize one particular accretion event in the simple form of a spherical cloudlet with a given impact parameter. We expect a higher impact parameter to lead to a more extended disk, and a smaller value to a smaller disk but also smaller misalignment due to the specific angular momentum of the cloudlet gas. However, the complexity of a more detailed study of late infall in the interstellar medium (ISM) or a parameter study is beyond the scope of this Letter, so that we do not aim to reproduce all facets of the IRAS~04125+2902 system. Additionally, interaction with the cloudlet gas may affect the giant planet's orbit, because the accreted primary disk mass is larger than the planet mass, $M_\mathrm{d,primary}=\SI{2.2}{\jupitermass}$. However, we do not consider the implications of this interaction, which has a significant impact in a flyby scenario \citep{nealon2025}, as it is outside the scope of this Letter.

Nevertheless, we considered the impact of the grid resolution on our results. We chose to initialize the simulation such that the cloudlet orbit is in the midplane of the spherical grid, and the companions orbit such that the initial misalignment between the two orbits is \SI{60}{\degree}. Grid-related forcing effects, though they are reduced in a spherical grid, promote the retention of the cloudlet in the midplane (e.g., \citealt{hopkins2015}). As a result, our setup promotes the misalignment we find for numerical reasons. We therefore run an identical simulation with a modified grid resolution of $N_r=140$, $N_\theta=440$, and $N_\phi=140$. This corresponds to 8 cps in colatitude at \SI{10}{\astronomicalunit}, which can alleviate this effect \citep{kimmig2024}, at the cost of a loss of resolution by a factor 2 in the other dimensions. We find $\bar{\vartheta}_\mathrm{d,o}=\SI{53.0}{\degree}$, which remains consistent with the observation within $1\sigma$. This test confirms that the found misalignment is not due to numerical effects. However, the disk is considerably less warped, indicating that the employed resolution is not sufficient to make quantitative statements. Details are discussed in Appendix \ref{sec:app_colatitude}.

\subsection{Comparison with observation}
The primary goal of this study is to show a possible pathway to the observed orbital misalignment of the transition disk around IRAS~04125+2902. While such a misalignment is indeed a natural outcome of late accretion as shown in our simulation, other observables of the system cannot be addressed with this setup.

The disk observed around the primary is a transition disk. The continuum dust emission shows an inner cavity extending to ${\sim}\SI{20}{\astronomicalunit}$, and the outer disk extends out to ${\sim}\SI{40}{\astronomicalunit}$. The gas disk appears more extended, with CO emission reaching out to ${\sim}\SI{120}{\astronomicalunit}$ (see Appendix \ref{sec:app_alma}). In contrast, our simulation, which does not include dust, shows a gas disk out to ${\sim}\SI{300}{\astronomicalunit}$. There is no inner cavity, though the inner edge of the simulation grid is located at \SI{5}{\astronomicalunit}. Even though we neglect the giant planet, it is unlikely to be the cause due to its low mass and semi-major axis. Even at higher mass and semi-major axis, the planet would either be too inclined to carve a gap \citep{xgp13}, or be subject to inclination damping inconsistent with the observations \citep{bk11}. Therefore, another mechanism must be responsible for its formation. While we find a warped disk (see Fig. \ref{fig:figure_3}), we stress that this result, especially the warp strength, is uncertain due to insufficient resolution.

We reproduce the present-day disk mass, but our simulated disk likely represents an early stage subject to subsequent evolution. The observed disk may have lost mass via viscous evolution, photoevaporation, and accretion. Dust growth and drift may influence the continuum emission. Being an earlier stage, our disk should have a higher mass than the present-day value. However, the final disk mass is a function of the initial cloudlet mass, which is a free parameter in our setup and could be increased accordingly. We note that an expanding cloudlet is a considerable simplification of ISM conditions, where mass is lost during the expansion and resulting outflow at the boundaries. However, this is considered in our choice of the initial cloudlet mass. In fact, the extended density structure that is the result of the full engulfment of the system in cloudlet gas during the initial encounter could be seen as representative of Bondi-Hoyle accretion in a filamentary structure, which is likely the dominant type of environmental interaction \citep{winter2024,pelkonen2025}

Lastly, the proposed late accretion event is not the only possible explanation for the observed system configuration. A stellar flyby could cause a misalignment, though it would likely perturb the rest of the system \citep{nealon2025}. Future work to confirm the position angle of the binary and planetary orbits with high precision will help distinguish between scenarios. Speculatively, the giant planet could also form in the accreted warped disk and become misaligned relative to the outer disk regions if formed in the inner disk regions for sufficient warp strength. However, planet formation in a warped disk is not a well-studied topic, and we do not explore this idea further in this Letter.

\subsection{A circum-secondary disk}
Our simulation reveals an additional, previously unexplored, feature of the system: a disk around the secondary. The accretion of mass around this star is inevitable if a late infall event is responsible for the misaligned disk around the primary. Observationally, its existence is unknown. Previous ALMA observations were not suitable to detect it. First, their sensitivity was too shallow to detect the circum-secondary disk, which is less massive than the circum-primary disk and accretes onto the secondary on a shorter timescale. Second, the maximum recoverable scale of the chosen configuration was only ${\sim}2.1^{\prime\prime}$, smaller than the binary separation, so that extended emission like a connection arc would have been missed. A deeper observation with a higher maximum recoverable scale, aimed at discovering this disk, could shed important light on the plausibility of a late infall event causing the observed disk-planet misalignment, offering broader insights into protoplanetary disk evolution and planet formation.

We find that the orientation of this disk is different from the primary disk, that it is evolving more quickly over time, and that it is connected to the primary through an arc of material with episodic mass transfer events. This is due to the mass ratio of the binary stars, $M_\mathrm{s}/M_\star=0.24$. For the less massive secondary, the instantaneous rest frame changes more rapidly with orbital phase due to the larger semi-major axis around the barycenter. Therefore, the angular momentum of the cloudlet relative to the secondary star, determining the disk orientation, evolves more strongly, and differs from the circum-primary disk. The simulation does not resolve the circum-secondary disk in detail, but it could serve as a test bed for future studies on circumstellar disks heavily influenced by binary interactions and their environment.

\section{Conclusion}\label{sec:conclusion}
In this Letter, we propose that the existence of a transition disk around IRAS~04125+2902 that is misaligned with respect to the other edge-on system components could be explained by the accretion of a gas cloudlet whose orbital plane is misaligned with respect to the system's original angular momentum. A 3D hydrodynamical simulation with a simplistic setup can reproduce relevant aspects of the system's configuration, misalignment and mass, despite the gravitational interaction with the secondary star. While a perfect match is not expected with our setup, and more detailed studies of the parameter space and the subsequent disk evolution after the accretion event may lead to a closer match. Indeed, the formation of second-generation protoplanetary disks with random orientation is a natural outcome of late-accretion onto a system and does not perturb the configuration of existing planets or binary orbits.

Our work serves as a proof of concept demonstrating that material inflowing from the larger-scale environment is imperative to consider when investigating disk-planet misalignment, and protoplanetary disk evolution in general.

\begin{acknowledgements}
We thank the anonymous referee for their thoughtful comments that helped greatly improved the manuscript. L.-A. H. acknowledges funding by the DFG via the Heidelberg Cluster of Excellence STRUCTURES in the framework of Germany's Excellence Strategy (grant EXC-2181/1 -- 390900948). The authors acknowledge support by the High Performance and Cloud Computing Group at the Zentrum für Datenverarbeitung of the University of Tübingen, the state of Baden-Württemberg through bwHPC and the German Research Foundation (DFG) through grants INST 35/1134-1 FUGG, 35/1597-1 FUGG, 37/935-1 FUGG and INST 37/1159-1 FUGG.
\end{acknowledgements}
\bibliography{references}
\begin{appendix}
\onecolumn
\section{ALMA observation of the circumstellar disk around IRAS~04125+2902}\label{sec:app_alma}
\begin{figure}[htp]
    \centering\includegraphics[width=.45\linewidth]{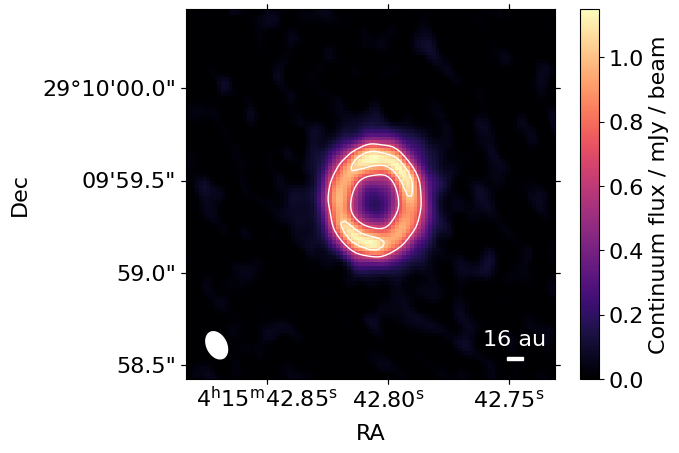}
    \includegraphics[width=.45\linewidth]{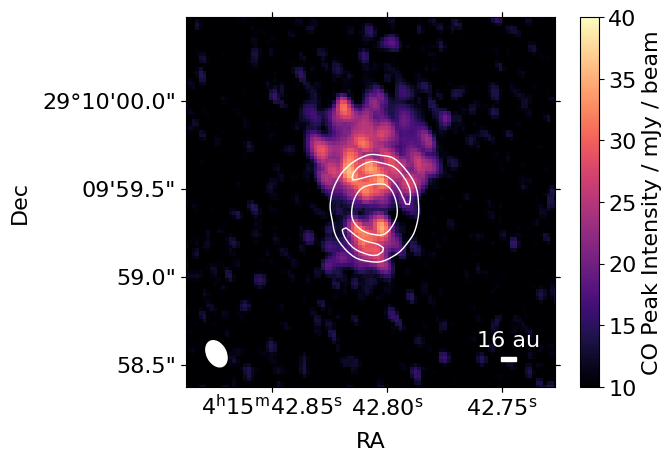}\\
    \includegraphics[width=.45\linewidth]{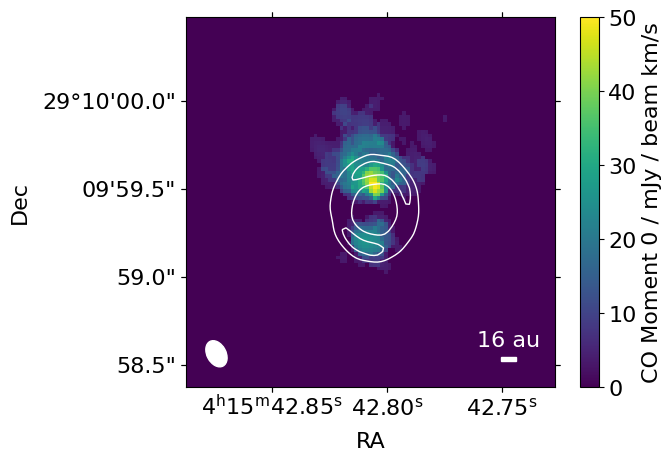}
    \includegraphics[width=.45\linewidth]{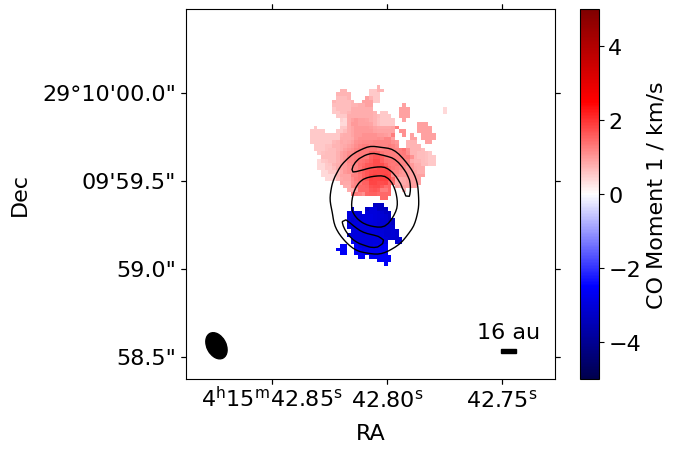}
    \caption{ALMA Band 6 observations of the circumstellar disk around the primary. The contour lines show the $10\sigma$ and $15\sigma$ continuum emission. The top left panel shows the \SI{1.3}{\milli\meter} continuum emission map, the top right panel the CO peak intensity map, the bottom left panel the CO moment 0 map, and the bottom right panel the CO moment 1 map.}
    \label{fig:alma}
\end{figure}
To determine the extent of the gas disk around the primary star, as well as to compare the disk arising in the simulation to the highest resolution images available, we show the ALMA observation of IRAS~04125+2902 \citep[2022.1.01302.S, PI: G. Mulders,][]{bosschaart2025,shoshi2025} in Fig. \ref{fig:alma}. A clear transition disk can be seen, with an inner cavity reaching out to ${\sim}\SI{20}{\astronomicalunit}$ and an outer dust disk out to ${\sim}\SI{40}{\astronomicalunit}$. This data confirms the inclination of $i_d\approx\SI{30}{\degree}$ based on the low-resolution SMA observation \citep{espaillat2015}. Using the peak intensity of the CO emission, we find that the gas component extends out to ${\sim}\SI{120}{\astronomicalunit}$.

\FloatBarrier\newpage
\section{Mass of second-generation disks}\label{sec:app_disk_mass}
\begin{figure}[htp]
    \centering\includegraphics[width=.4\linewidth]{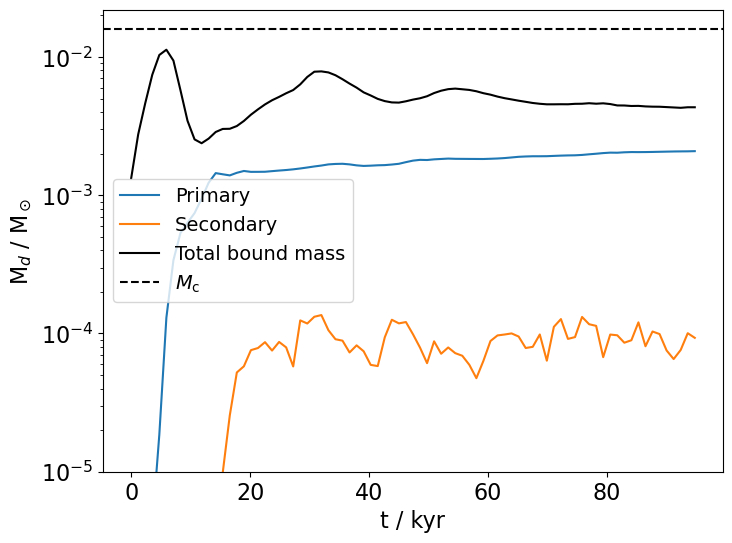}
    \caption{Total masses of the primary (blue line) and secondary (orange line) disks, as well as the initial cloudlet mass $M_\mathrm{c}$ (black dashed line) and total mass of bound gas (black solid line).}
    \label{fig:figure_2}
\end{figure}%
The total gas mass of the disk around the primary is different from the initial cloudlet mass. Leftover gas with a total mass of $M_\mathrm{rest}=\SI{2.2e-3}{\solarmass}$ is bound to the binary system at the end of the simulation, subject to eventual accretion in subsequent evolution. A significant fraction of gas is lost as a result of the cloudlet expansion, $M_\mathrm{lost}=\SI{1.2e-2}{\solarmass}$, implying an accretion efficiency of 27\%. Some accreted material also forms a circum-secondary disk. The time evolution of the gas disk masses around both stars and the total mass of gas bound to the binary system is shown in Fig. \ref{fig:figure_2}. Gas bound to the secondary, $v^2/2\leq GM_\mathrm{s}/\left|\vec r-\vec r_s\right|$ and within its Hill radius $r_H$ was considered part of its disk. Here, $v$ is the speed of the gas parcel, and $\vec r$ and $\vec r_s$ are the location vectors of the gas parcel and the secondary, respectively. Mass bound to the primary and inside a sphere with radius $a_\mathrm{bin}$, with $a_\mathrm{bin}$ the binary orbit semi-major axis, around the origin was considered as part of the primary disk.

The steady increase in disk mass over time is related to the fall-back of mass left over from the initial accretion event, whereas oscillations of the disk masses indicate episodic mass transfer between the two disks. Both effects contribute to the overall disk orientation and its warp. At the end of the simulation, the primary disk mass is ${\sim}\SI{2.1e-3}{\solarmass}$, consistent with values obtained using millimeter flux \citep{barber2024}. The ratio of disk masses is $M_\mathrm{d,secondary}/M_\mathrm{d,primary}\sim 4.5\%$.

\FloatBarrier\newpage
\section{Time evolution of the simulation}\label{sec:app_time_series}
\begin{figure}[htp]
    \centering\includegraphics[width=\linewidth]{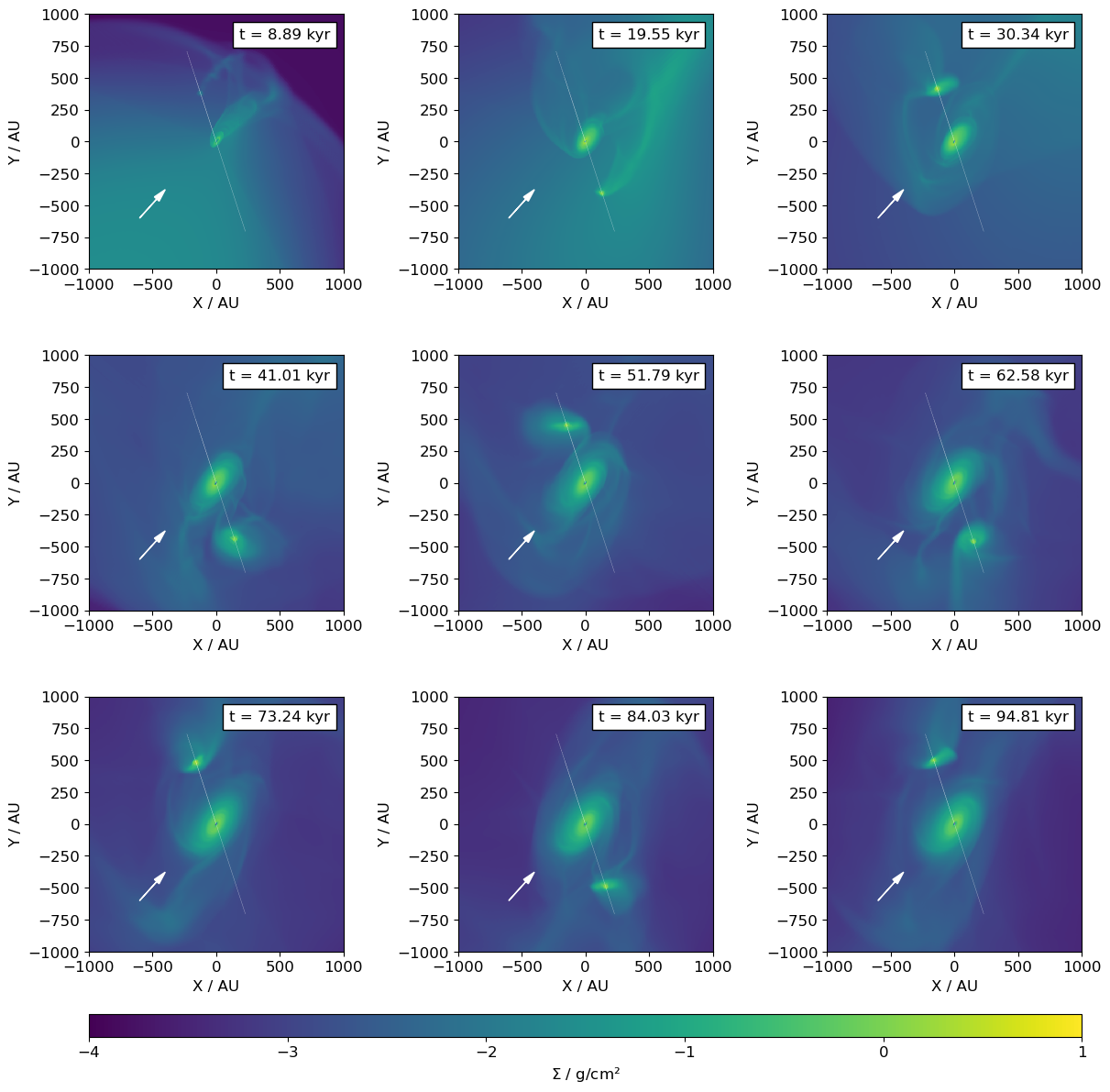}
    \caption{Time series of the gas column density found in the simulation. The camera perspective is chosen as in Fig. \ref{fig:figure_1}. The white arrows denote the initial infall direction, and the white dashed line represents the orbit of the secondary star.}
    \label{fig:time_series}
\end{figure}
Figure \ref{fig:time_series} shows nine snapshots of the gas column density found in the simulation. It can be seen that both the primary and secondary star accrete matter from the encounter with the cloud, forming disks with different orientations, though the resolution is not high enough to fully resolve the secondary disk. Even though the secondary regularly perturbs the primary disk while it is still accreting left-over material, interactions become less severe over time, eventually resulting in the final, misaligned disk.

\FloatBarrier\newpage
\section{Proposed system formation scenario}\label{sec:app_formation_scenario}
\begin{figure}[htp]
    \centering\includegraphics[width=.9\linewidth]{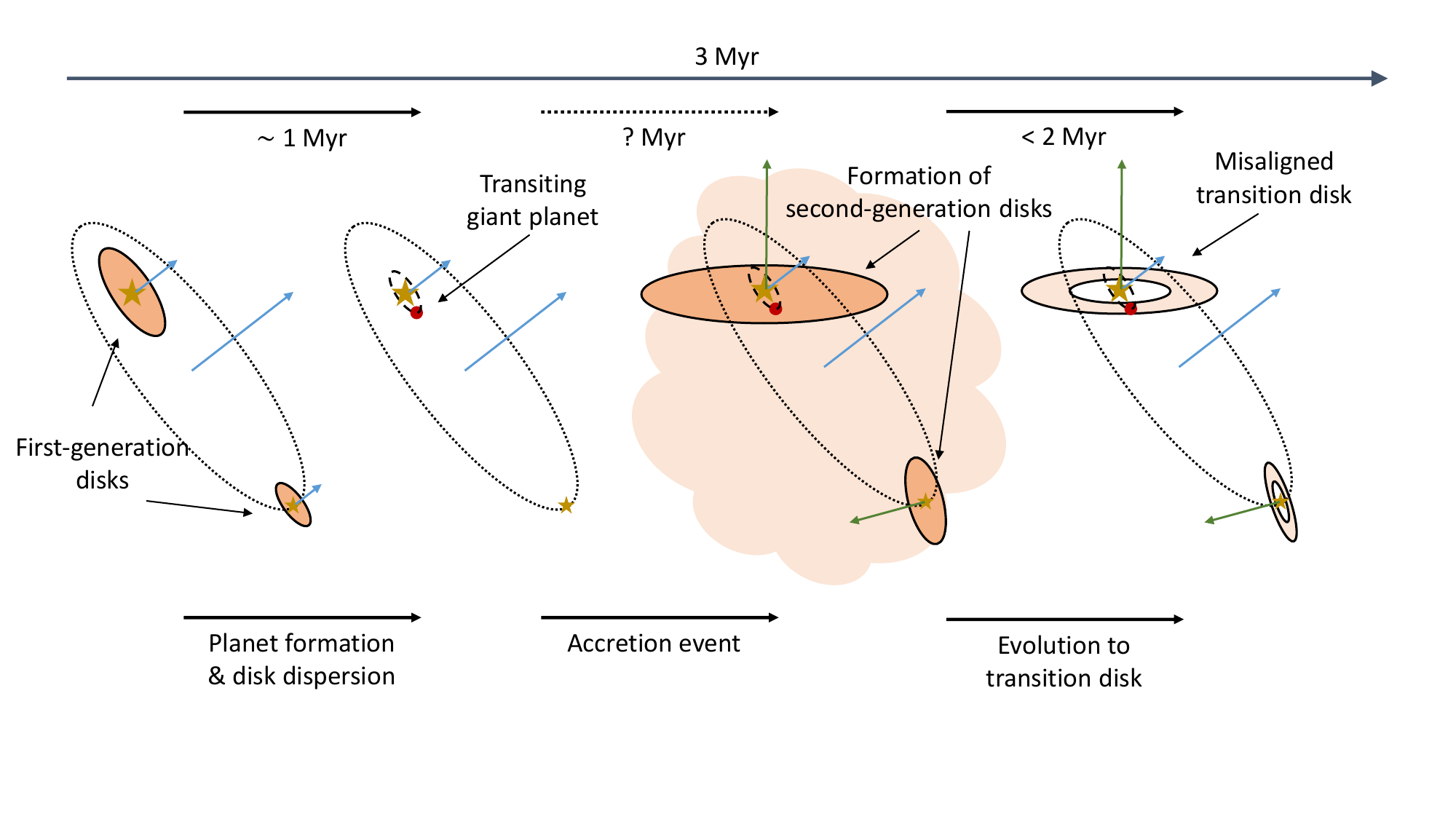}\vspace*{-3em}
    \caption{Formation and evolution of the IRAS~04125+2902 and companion binary system in the scenario that a late infall event is the cause for the observed misaligned transition disk. The various evolutionary stages over time are depicted from left to right. The blue arrows denote the original angular momentum vector orientation of the natal cloud and the resulting primordial system. Dashed and dotted black lines represent the orbits of the giant planet and companion star. Orange colored areas represent gas in the form of protoplanetary disks or a cloudlet. The green arrows indicate one representation of the random alignment of the second-generation disks expected from the accretion event. The scenario is exaggerated; in reality, at least some material from the first-generation disk will likely be leftover during the accretion event.}
    \label{fig:appendix_sketch}
\end{figure}%
\noindent Based on the late infall scenario presented in this Letter, we show a potential full system evolutionary scenario as a sketch in Fig. \ref{fig:appendix_sketch}. Initially, the system was created from a natal molecular cloud, inheriting the angular momentum from it and thereby forming the primordial protoplanetary disks around both stars in an aligned, edge-on configuration. The transiting giant planet observed via transit events forms in the disk around the primary star, while the disk evolves and loses mass through mechanisms like accretion and photoevaporative or magnetic winds.

Subsequently, the system encounters a gas cloud or, more generally, an over-density caused by ISM turbulence. Observations of the Taurus star-forming region \citep{garufi2024} may indicate that such environmental interactions are ubiquitous, making it possible for a cloudlet encounter to occur at the required time. As a result, two new, second-generation disks are accreted onto the two stars, whose first-generation disks may have already lost a considerable amount of mass depending on the timing of the encounter and the mass loss rate. Their orientation is random compared to the primordial configuration as the material's origin is unrelated to the natal cloud. It may not be necessary for the first-generation disks to have lost the majority of its mass at this point; a cloudlet encounter with an existing primordial disk could lead to a broken disk where each component maintains its alignment \citep{kuffmeier2021}, and the inner disk could continue to dissipate after the encounter, eventually resulting in the observed present-day inner cavity. In fact, a leftover primordial disk would aid the formation of the inner cavity, which would otherwise require processes like photoevaporation operating on longer timescales.

Previous work suggests that planetesimals, acting as the seeds for planets to grow by pebble accretion, form early and quickly \citep{lenz2020,dd18}. In that case, a planet could grow to a mass of \SI{0.3}{\jupitermass} during ${<}\SI{1}{\mega\year}$ \citep{bitsch2015}, depending on parameters like the location of the seed and the pebble flux. Observations indicate that ${\sim}30\%$ of disks may have dissipated after \SI{1}{\mega\year} \citep{mamajek2009}. If the disk has not fully dissipated, a viscous or magneto-hydrodynamical disk wind $\alpha_\mathrm{\{SS,DW\}}=\num{2e-3}$ would be required to clear a \SI{20}{\astronomicalunit} cavity in \SI{2}{\mega\year}. Here, we assume that the disk is broken with no supply from the outer disk. For the proposed chain of events to be feasible, early planetesimal formation and sufficient solid availability is required. Therefore, the scenario is favored if planet formation starts early, that is, before the Class II evolutionary stage of the disk. Furthermore, disks at this stage are more compact \citep{lebreuilly2024}, so that the cavity could indeed be a result of a broken inner disk.

Additionally, the two second-generation disks are misaligned with respect to each other due to the orbital motion of the secondary and large physical separation between the two stars. The accretion event does not perturb the alignment of the binary orbit, which differentiates this scenario from a gravitational interaction with a third star. To arrive at the transition disk that is observed today, further disk evolution may take place, which does not affect the disks' orientation, so that the misalignment caused by the accretion is kept.

\FloatBarrier\newpage
\section{Simulation focused on high colatitude resolution}\label{sec:app_colatitude}
\begin{figure}[htp]
    \centering\includegraphics[width=.45\linewidth]{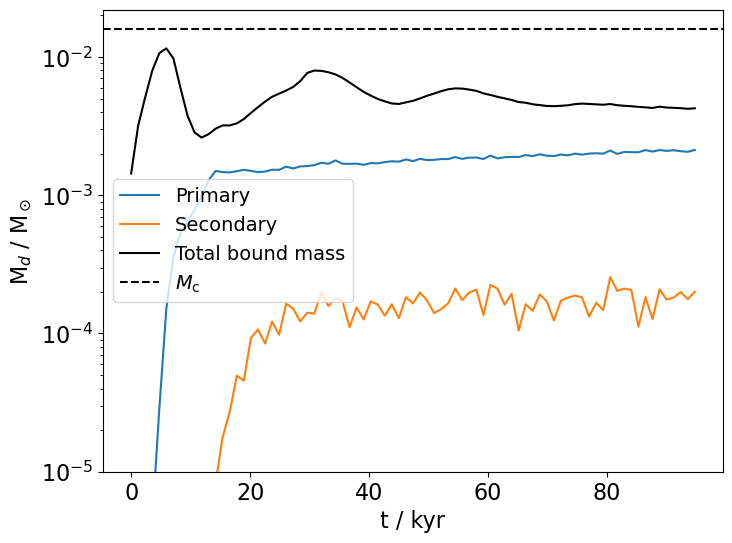}\hfill
    \includegraphics[width=.48\linewidth]{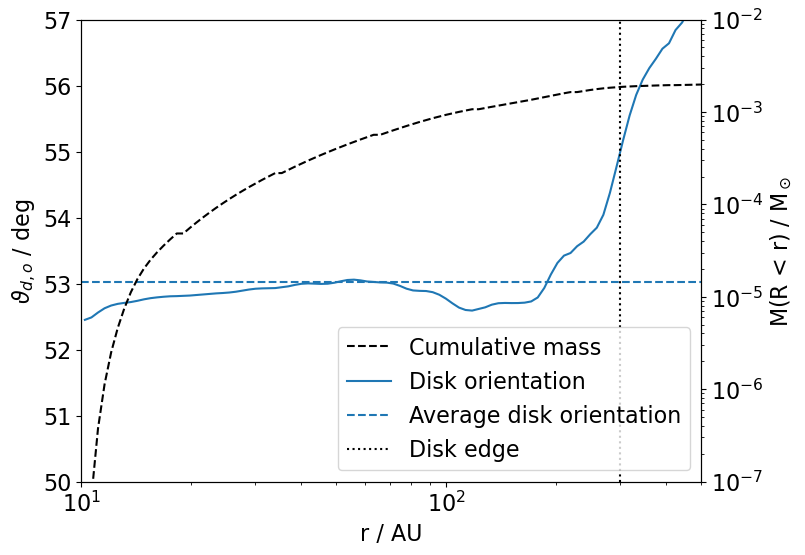}
    \caption{Disk and total bound mass time evolution and primary disk alignment resulting from the simulation focusing on high colatitude resolution. Left: As Fig. \ref{fig:figure_2}. Right: As Fig. \ref{fig:figure_3}, but with an adapted vertical axis range.}
\label{fig:colatitude}
\end{figure}

We performed a simulation whose setup is identical to the one described in Section \ref{sec:methods}, but with changed resolution. In order to verify the robustness of the binary-disk misalignment against grid-related effect, we chose $N_r=140$, $N_\theta=440$, and $N_\phi=140$ to emphasize the resolution in colatitude. In doing so, we achieve 8 cells per scale height at \SI{10}{\astronomicalunit} (12 at \SI{50}{\astronomicalunit}), or a resolution of $\Delta\phi = \SI{2.6}{\degree}$, $\Delta\theta=\SI{0.32}{\degree}$ and $\Delta r/r=0.044$. While the results are qualitatively similar, we find some differences between the simulations, shown in Fig. \ref{fig:colatitude}.

First, the distribution of material between primary and secondary disk at the end of the simulation is different. This is shown in the left panel of Fig. \ref{fig:colatitude}. Here, the secondary disk is more massive, $M_{d,s}=\SI{2e-4}{\solarmass}$. The resulting mass fraction between the two disks is $9.4\%$, which is a factor ${\sim}2$ larger than what we find for the simulation with balanced resolution. This value is slightly higher than estimated from the ratio of Bondi accretion rates, $\dot M_\mathrm{bondi,s}/\dot M_\mathrm{bondi,p} \approx 6\%$. The leftover mass is $M_\mathrm{rest}=\SI{1.9e-3}{\solarmass}$ here, implying the same accretion efficiency as in the simulation with balanced resolution, 27\%.

Second, the mass-weighted average of the disk orientation is lower than for the simulation with balanced resolution, $\bar{\vartheta}_\mathrm{d,o}=\SI{53.0}{\degree}$. Because this value is still consistent with the observed value within $1\sigma$, we conclude that our main result does not depend on the choice of the resolution. However, the radial profile of the orientation differs. As shown in the right panel of Fig. \ref{fig:colatitude}, the disk warp is significantly less pronounced. In particular, the lower misalignment between the binary orbit and the inner disk regions, as well as the steep increase in misalignment with radius, are not recovered.

\end{appendix}
\end{document}